# Thermally adaptive fabric inspired by Morpho butterfly for all-season comfort and visible aesthetics

Yan Wang[1]†, Zhuowen Xie[1]†, Wangkai Jiang[1,2]†[*], Ting-Ting Li[1], Honglei Cai[1], Jun Zhang[1], Hui Wang[1], Shuo Shi[2], Jintu Fan[2*], Jianchen Hu[1*], Ke-Qin Zhang[1*]

1 National Engineering Laboratory for Modern Silk, College of Textile and Clothing Engineering, Soochow University, Suzhou 215123, P. R. China.
2 School of Fashion and Textiles, The Hong Kong Polytechnic University, Kowloon, Hong Kong 999077, P. R. China.

†These authors contributed equally to this work.

*Corresponding authors. E-mails: wangkai.jiang@polyu.edu.hk (W. J.); jintu.fan@polyu.edu.hk (J. F.); hujianchen@suda.edu.cn (J. H.); kqzhang@suda.edu.cn (K.-Q. Z.)

**Abstract:** The longstanding challenge of transitioning from static, appearance-limited passive daytime radiative cooling (PDRC) materials to systems that are both dynamically adaptive and aesthetically versatile in personal thermal management stems from the inherent compromise between color saturation and cooling power. Inspired by the Morpho butterfly, which decouples structural color from thermal function, we introduce a modular adaptive textile that dissolves this conflict. A dynamic thermochromic membrane autonomously switches solar reflectance from ~0.6 (heating state) to ~0.9 (cooling state), while an APC layer supplies angle-independent, high-saturation structural color independent of the thermal state and fully replaceable, enabling on-demand color rewriting. Consequently, outdoor tests show 5–7 °C surface-temperature reduction versus commercial colored fabrics under hot conditions, with no compromise in solar-heating performance under cold conditions. Complemented by hydrophobicity and breathability, this color-reconfigurable, energy-free approach offers a general pathway toward aesthetic and adaptive thermal comfort.

## 1 Introduction

The pursuit of energy-efficient personal thermal management (PTM)(*1-3*) is crucial amid surging global energy demands, driven significantly by the need to manage the vast and diverse solar energy(*4-6*) incident on Earth (Fig. 1A). Passive daytime radiative cooling (PDRC) materials,(*7-9*) leveraging the atmospheric transparency window (8-13 μm) for heat dissipation and reflecting a substantial portion of incident solar radiation, offer a key zero-energy cooling solution. However, a critical bottleneck emerges when integrating vibrant color: since visible (VIS) light constitutes nearly half of the solar energy spectrum (Fig. 1B), high-saturation colors often rely on absorption in this region, thereby diminishing the material's solar reflectance and counteracting its cooling potential.(*10*) This inherent trade-off, coupled with the predominantly static nature of current colored PDRC materials, limits their adaptive capacity and broader applicability.

To overcome these limitations and achieve dynamic thermal regulation, a fundamental shift towards actively managing solar absorption across the VIS spectrum is necessary.(*11-14*) We propose a conceptual framework centered on a dynamic optical switch (Fig. 1C), wherein a material toggles between a high-reflectance state for cooling under intense solar irradiation and a controlled absorption state for harnessing thermal energy when needed. Theoretical modeling based on this principle (Fig. 1D and fig. S1) indicates that such bidirectional regulation of VIS light can yield substantial modulation of radiative heat transfer, enabling significant cooling and heating capabilities solely through VIS spectrum management. For example, Zhu et al.(*15*) fabricated a fabric functionalized with thermochromic microcapsules (TCMs) that were encapsulated in graphene and barium sulfate coatings, adapting to ambient temperature ($T_{amb}$) fluctuations and shifting color between black and white. This fabric exploited optical switching to maximize temperature regulation. However, such approaches often face challenges in decoupling color from thermal management, limiting color diversity. This concept finds compelling precedent in nature. While prior studies have established that structurally identical coloration on different substrates yields divergent optical properties, its thermal implication remains unexplored (fig. S2A).(*16-18*) We observed this phenomenon directly in Morpho butterfly wings under simulated solar irradiation. Remarkably, the dorsal structural color layer exhibits distinct appearances over the ventrally located black versus white substrates: the black-backed regions manifest as blue with significant photothermal heating, whereas the white-backed regions appear whitish with negligible heat gain (Fig. 1E and fig. S2). This occurs because the white substrate reflects the transmitted complementary light (yellow) back through the structure, preventing heat accumulation, while the black substrate absorbs it. This natural stratification emphasizes a critical pathway: by isolating the color-producing function from the thermal-responsive substrate, dynamic thermal management can be decoupled from static visual appearance.

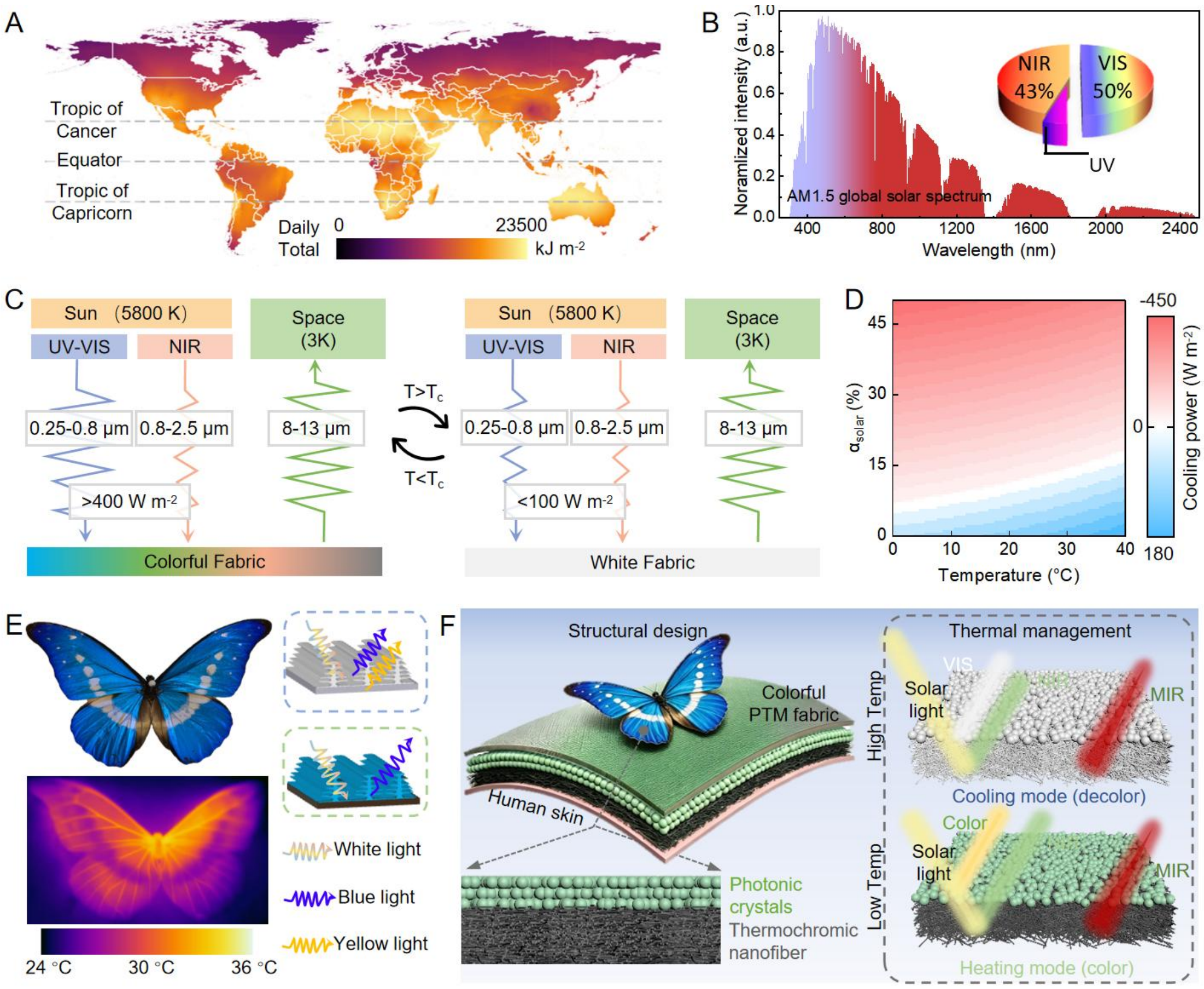

**Fig. 1.** Design of CSA fabric and conceptual framework. (**A**) Global spatial distribution of solar radiation. (**B**) Spectral characteristics of the AM1.5 global solar spectrum, insert: proportional distribution of solar energy. (**C**) Switching mechanism of optical and thermal radiation properties in thermochromic passive daytime radiative cooling (PDRC) materials at different power densities of the absorbed VIS light and ambient temperatures. (T: ambient temperature, $T_c$: phase-change temperature of the fabric). (**D**) Theoretical simulations of the relationship between solar absorbance and thermal power (positive: cooling power; negative: heating power). (**E**) Digital and infrared thermal photographs of a pair of Morpho butterfly wings (the dorsal side of the Morpho butterfly wings) under simulated solar irradiation (600 W m$^{-2}$). (**F**) Structural design and thermal management mechanism of the colorful smart adaptive (CSA) fabric (mimicking the structure of Morpho butterfly wings for the biomimetic design).

Inspired by this bio-insight, we engineered a colorful smart adaptive (CSA) fabric featuring a thermally responsive optical architecture. As schematically depicted in Fig. 1F, our design integrates a thermochromic nanofiber membrane with a structurally colored amorphous photonic crystal (APC) layer. The membrane reversibly switches its optical properties between a reflective white state and an absorptive dark state, thereby effectively modulating solar absorption and reflectance. The APC layer provides stable, high-saturation color independent of the underlying thermal state, while the nanofiber matrix ensures high mid-infrared emissivity ($\varepsilon_{MIR}$) (fig. S3). This synergistic combination allows the fabric to autonomously regulate its energy exchange with the environment: maintaining high solar reflectance ($\rho_{solar}$≈0.9) and cooling in high thermal load conditions, and enhancing solar absorption ($\alpha_{solar}$≈0.4-0.5) for passive heating when needed. Notably, it achieves a maximum cooling effect of 5-7 °C under high temperatures while maintaining solar-heating performance within <1 °C of commercial reference fabrics under low temperatures, which means the black heating in winter and the white cooling in summer from a single textile. Furthermore, it allows for the recoloring of thermal management fabrics, promising to meet varied aesthetic demands that are often sacrificed in traditional designs. Such decoupled structures can facilitate easy color customization without compromising cooling or heating efficiency, addressing the long-standing issue of color-thermal integration.

## 2 Results

### 2.1 Characterization and optical properties of SA fabric

To realize intelligent temperature regulation, we fabricated a smart adaptive (SA) fabric *via* electrospinning poly(vinylidene fluoride-co-hexafluoropropylene) (PVDF-HFP) nanofiber membranes embedded with TCMs (fig. S4). The incorporated TCMs enable reversible switching of the membrane's optical appearance between a black (heating state) and a white (cooling state).(*14*) Scanning electron microscopy (SEM) (Fig. 2A) reveals a randomly interwoven nanofibrous network, in which micron-sized TCMs are uniformly embedded within the inter-fiber spaces and partially wrapped by the surrounding fibers. To optimize radiative cooling, the fiber diameter distribution was statistically analyzed based on Mie scattering theory,(*19, 20*) as diameter critically governs scattering behavior and thus solar reflectance. The measured fiber diameters exhibit a broad distribution ranging from 0.1 to 1.3 μm, with the distribution centered at approximately 0.68 μm (Fig. 2B). Further MATLAB-based(*21*) Mie scattering analysis of the scattering efficiency ($Q_{sca}$) as a function of fiber diameter (Fig. 2C) indicates that this broad diameter distribution enables efficient scattering across the solar spectrum of 0.25-2.5 μm. Specifically, fibers with diameters >680 nm efficiently scatter near-infrared (NIR, 0.75-2.5 μm) light, while those <680 nm contribute to VIS light (0.4-0.75 μm) scattering. Consequently, the SA fabric maintains high NIR reflectance (providing a foundation for cooling mode) and enhanced total solar reflectance in the cooling state, where TCMs do not absorb VIS light.

The mechanism (Fig. 2D) of reversible thermochromism relies on TCMs composed of a color former (incorporating lactone and fluorescein dyes), and a color developer.(*22*) At lower temperatures (<25 °C, heating state), interaction between the color former and developer induces lactone ring-opening, forming an extended conjugated system that absorbs VIS light (black state). At elevated temperatures (>25 °C, cooling state), the color former undergoes molecular rearrangement, not merely electron gain, leading to lactone ring closure, disrupting the conjugated structure. This transition shifts the TCMs from broadband VIS absorption (black) to high reflectance (white), as confirmed by macroscopic optical changes. Thickness optimization was performed to maximize the solar-modulation capability of the SA fabric. As shown in fig. S5, increasing its thickness from 0.25 to 0.48 mm enhanced the solar reflectance in both temperature states, and the 0.48 mm sample exhibited the largest difference in solar reflectance between the heating and cooling modes.

Additionally, the Fourier transform infrared (FTIR) spectra confirm that the SA fabric exhibits distinct characteristic absorption bands such as the C-F vibration of PVDF-HFP (as shown in fig. S6) in the atmospheric transparency window (8-13 μm), with high $\varepsilon_{MIR}$ (≈0.94, Fig. 2E) maintained in both states. This combination of switchable solar reflectance and constant high thermal emissivity underpins bidirectional thermal management. To demonstrate intuitive thermal visualization, the SA fabric was placed on adjacent temperature-controlled plates set to 15 °C (low-temperature end) and 35 °C (high-temperature end), exhibiting a clear visual temperature difference (fig. S7). The full spectral response of the fabric is summarized in Fig. 2E, validating its dual-mode optical regulation.

The reversible optical switching of the SA fabric exhibits excellent stability. Cyclic VIS reflectance measurements (Fig. 2F and fig. S8) showed minimal degradation after 10 thermally induced switching cycles (15-35 °C), confirming essential stability for practical applications. To validate the temperature-adaptive thermal management of the SA fabric, we conducted indoor simulated sunlight tests (Fig. 2G) using a solar simulator (0-1000 W $m^{-2}$, AM1.5G filter) under controlled conditions ($T_{amb}$=10 °C, no forced convection), with samples including SA fabric, commercial black fabric (traditional solar absorber), and white radiative cooling (RC) fabric (pristine electrospun PVDF-HFP membrane) as the reference. As light intensity increased from 0 to 1000 W $m^{-2}$ gradually, the SA fabric (Fig. 2H) initially operated in a heating state: below 25 °C (TCMs in absorptive black state, no VIS color change), its temperature evolution mirrored that of commercial black fabric, exhibiting a similar rapid heating rate. At ~400 W $m^{-2}$, the SA fabric's temperature surpassed 25 °C, triggering TCMs to switch to a reflective white state (cooling mode). Thereafter, even as light intensity further increased to 1000 W $m^{-2}$, its heating rate decelerated significantly, stabilizing at ~28 °C. In contrast, commercial black fabric continued heating to a temperature of ~45 °C at 1000 W $m^{-2}$, while the white RC fabric maintained a consistently low temperature (~18 °C) across all intensities. These results confirm the SA fabric's autonomous regulation: it mimics black cotton for efficient passive heating at low temperatures (<25 °C) and switches to white-like cooling at high temperatures (>25 °C), limiting temperature rise to ~28 °C even under intense sunlight, thereby demonstrating robust temperature-adaptive thermal management.

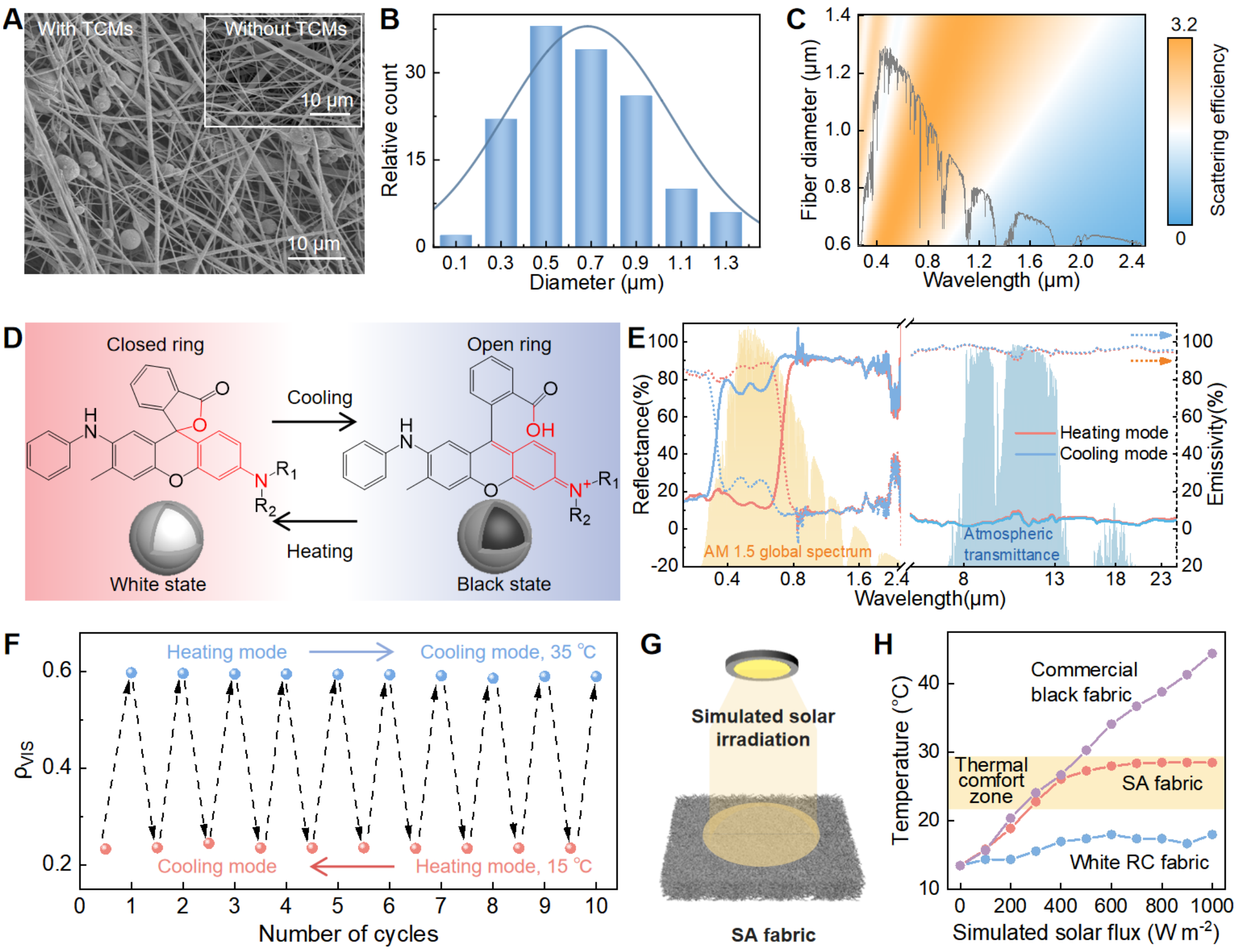

**Fig. 2.** Characterization and optical-thermal properties of SA fabric. (**A**) SEM images of the electrospun PVDF-HFP nanofiber membrane with TCMs (inset: membrane without TCMs). (**B**) Statistical distribution of fiber diameters. (**C**) MATLAB-calculated $Q_{sca}$ versus fiber diameter based on Mie scattering theory. (**D**) Mechanism of reversible thermochromism in TCMs (color former, developer). (**E**) Full spectral response of the SA fabric in both cooling and heating modes. (**F**) Cyclic stability of VIS reflectance after 10 thermally induced switching cycles (15-35 °C). (**G**) Illustration of the indoor experiment simulating sunlight intensity. (**H**) Temperature evolution of SA fabric, commercial black fabric, and white RC fabric reference under increasing light intensity (0-1000 W $m^{-2}$).

## 2.2 Optical-thermal performance of CSA fabric

To achieve multi-color integration in thermal management fabrics while decoupling aesthetics from functionality, we engineered a CSA fabric by integrating colloidal APCs onto the surface of the previously developed SA fabric (Fig. 3A). This layered architecture combines the SA fabrics' thermochromic switching (heating/cooling states) with the APCs' angle-independent structural color, enabling customizable coloration without compromising thermal regulation. The CSA fabric features a hierarchical structure: a bottom layer of the SA fabric (PVDF-HFP nanofiber membrane embedded with TCMs) for reversible solar absorption/reflection, and a top layer of colloidal APCs deposited by spray-coating after synthesis by emulsion polymerization (details in figs. S9-S13). The cross-sectional microstructure of the CSA fabrics (including APCs and nanofiber matrix with TCMs), and the surface morphology of each part (APCs and nanofiber matrix with TCMs) were investigated (Fig. 3B), which reveals that the upper layer consists of APCs and the lower layer consists of a nanofiber matrix with uniformly distributed TCMs. The confined assembly of APCs on the nanofiber surface minimizes structural defects, endowing the fabric with high-brightness, highly saturated structural colors.

The structural colors of the CSA fabrics originate from wavelength-selective reflection/scattering within the APC layer composed of PS nanospheres.(*23, 24*) The APC layer possesses short-range order but lacks long-range periodicity, which enables vivid structural colors while suppressing strong angle-dependent color shifts. By adjusting the diameter of the PS nanospheres, the reflection peak can be tuned across the visible region. As shown in fig. S11, PS nanospheres with average diameters of 178, 215, and 240 nm generate blue, green, and red structural colors, respectively. The corresponding color coordinates are summarized in the CIE 1931 chromaticity diagram in Fig. 3C, confirming the tunable coloration of the CSA fabrics. In addition, the angle-dependent spectra in fig. S13 show that the APC-based CSA fabric maintains a relatively stable reflection band at different observation angles, demonstrating its non-iridescent structural color.

The thermal regulation of the CSA fabric hinges on the reversible switching of $\rho_{solar}$ between two modes: passive heating, where TCMs adopt an absorptive state, and radiative cooling, where TCMs switch to a reflective state. As illustrated in Fig. 3D, three representative CSA fabrics (blue, green, and red) exhibit distinct optical behaviors. In the heating state (solid lines), all fabrics display lower $\rho_{solar}$ (≈0.5-0.6, particularly in the VIS region), dominated by TCMs absorption to facilitate passive heating. Crucially, each color variant retains a reflection peak characteristic of structural color (from APCs) in both heating and cooling states, confirming the stability of the color-generating mechanism independent of thermal switching. In the cooling state (dashed lines), $\rho_{solar}$ rises sharply to ≈0.8-0.9, attributed to the white state of the substrate SA fabric and enhanced reflection from the APC layer. The reflectance difference between states exceeds 30% across the solar spectrum (0.25-2.5 μm), enabling efficient bidirectional thermal control. The integration of the APCs-based structural color layer does not compromise the SA fabric's infrared emissivity. Taking the green CSA fabric as an example (fig. S14), MIR emissivity analysis reveals a high $\varepsilon_{MIR}$≈0.94 within the atmospheric transparency window (8-13 μm) in both heating and cooling modes. This consistency holds for all structural colors, indicating the APC layer minimally interferes with the thermal radiation properties of the underlying substrate. Cyclic stability tests (20 thermally induced switching cycles between 15 °C and 35 °C, Fig. 3E and fig. S15) further validate the cyclic stability of the APC-integrated structure over thermal switching.

To demonstrate practical applicability, Fig. 3F shows a flower-patterned CSA fabric (spray-coated with PS nanospheres with diameters of 178, 215, 240 nm onto the SA substrate fabric, fig. S16) under simulated sunlight at two $T_{amb}$: cold (20 °C) and hot (35 °C) conditions. The patterned regions exhibit clear color transitions: at 35 °C, the colors fade gradually due to TCM switching to the reflective state, while at 20 °C, vibrant hues reappear as TCMs absorb light. Temperature mapping reveals that the colored CSA fabric (with black TCMs) reaches higher temperatures than the ambient under heating mode, whereas colorless CSA fabric remains relatively cooler. Critically, selective adjustment of the TCMs' phase transition temperature can precisely tune the transition interval between cooling and heating states and enable the fabric to exhibit corresponding colors (fig. S17). This visual-thermal correlation emphasizes the fabrics' ability to dynamically balance aesthetic expression and thermal function in real-world scenarios.

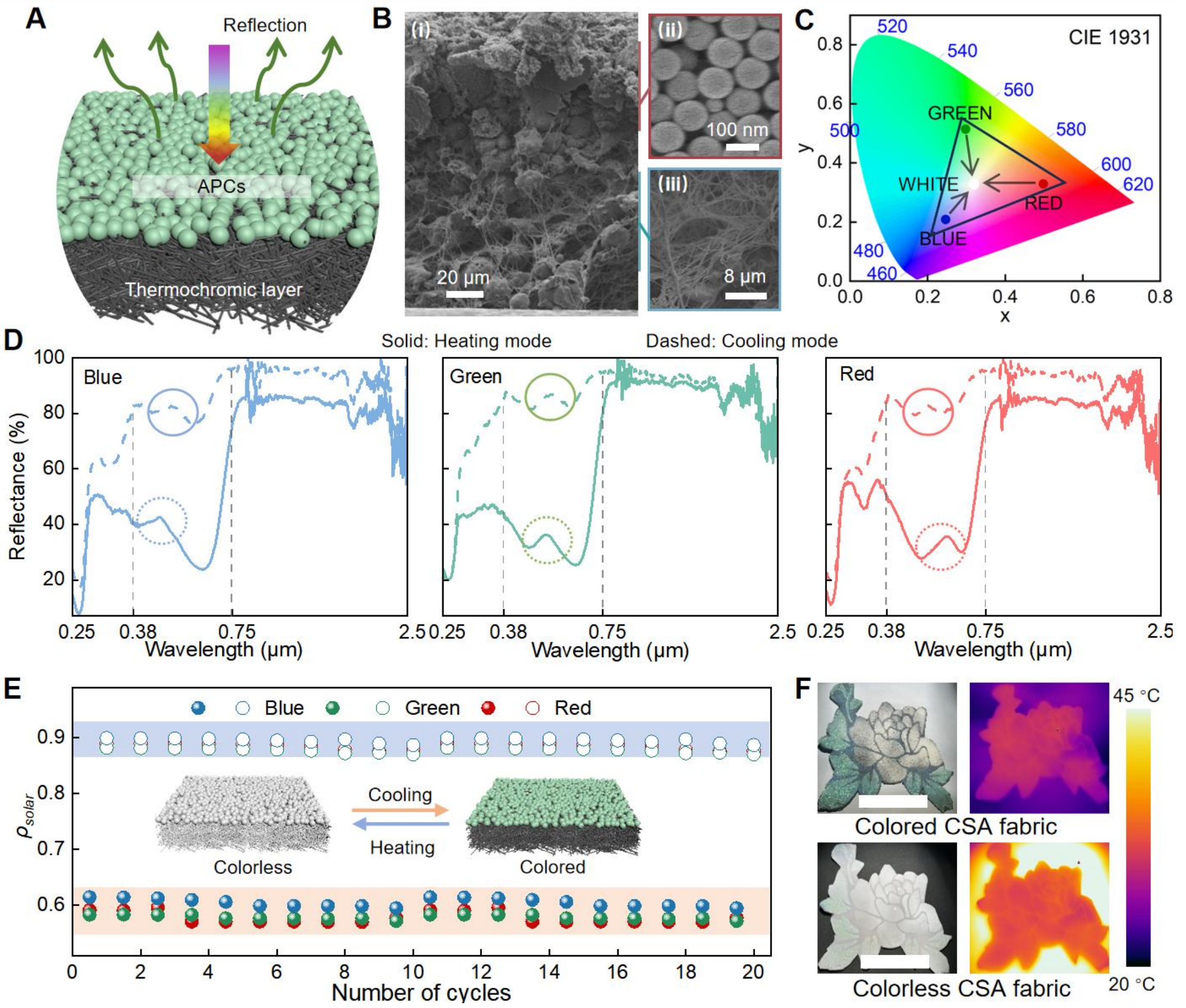

**Fig. 3.** Optical-thermal performance of CSA fabric. (**A**) Schematic of CSA fabric fabrication *via* spraying colloidal APCs onto SA fabric. (**B**) (i) Cross-sectional SEM image of CSA fabric (top: APCs; bottom: nanofiber matrix with TCMs), and the surface morphology SEM images of (ii) APCs and (iii) nanofiber matrix with TCMs. (**C**) CIE 1931 chromaticity diagram confirming structural colors saturation. (**D**) Spectra of blue, green, and red CSA fabrics in heating (solid) and cooling (dashed) states (Scale bar: 2 cm). (**E**) Cyclic stability of the thermal switching function over 20 cycles (15-35 °C). (**F**) Photos and infrared thermal images of flower-patterned CSA fabric under simulated sunlight (top: 20 °C, bottom: 35 °C, scale bar: 3 cm).

## 2.3 Outdoor evaluation of dual-mode CSA fabric

To validate the autonomous thermal regulation of the CSA fabric under real-world conditions, outdoor tests were conducted in Suzhou, Jiangsu Province of China (31.3° N, 120.7° E), over three hours under low (≈15 °C) and high (≈35 °C) temperature regimes. Fig. 4A documents photographic evidence of clear temperature-dependent color transitions. At elevated temperatures of approximately 35 °C, all CSA fabrics transitioned to a white state upon TCM activation, irrespective of their initial structural hues including blue, green, and red as well as the underlying black state of the thermochromic membrane. Conversely, at reduced temperatures near 15 °C, the fabrics retained vivid structural colors. These observations confirm that the APC-based structural coloration mechanism operates in concert with autonomous thermal regulation. The experimental setup (Fig. 4B and Fig. 4C) was designed based on a classic PDRC measurement platform(*25, 26*) to minimize parasitic heat exchange. It consisted of a thermally insulated chamber housing three circular CSA fabric samples (diameter: 90 mm). The chamber featured an aluminum foil-coated foam enclosure (to reflect external radiation and reduce conductive heat loss), a foam-insulated sample stage (to isolate samples from ground heat), and copper thermal contact sheets (to ensure uniform heat transfer between samples and temperature sensors). The sample temperature was measured directly using thermocouples, while the ambient temperature and solar irradiance were recorded by a weather station and a pyranometer, respectively. This configuration enabled precise quantification of the fabrics' thermal response under natural sunlight.

Under high-temperature conditions (humidity and wind speed were recorded in fig. S18A), Fig. 4D (left panel) shows that the surface temperatures of the CSA fabrics were consistently 5-7 °C (figs. S19A-C) lower than those of corresponding commercial-colored fabrics. This cooling effect is attributed to the CSA fabrics' transition to a reflective state, where TCMs switch to a white state, achieving high solar reflectance ($\rho_{solar}$≈0.8-0.9) and stable mid-infrared emission ($\varepsilon_{MIR}$≈0.94) for radiative heat dissipation. At low temperatures (humidity and wind speed were recorded in fig. S18B), the CSA fabrics retained their structural color (photographic evidence in Fig. 4A), with lower solar absorptance ($\alpha_{solar}$≈0.4-0.5) only slightly lower than that of commercial colorful fabrics (Fig. 4D, right panel). The steady-state temperature of the CSA fabrics differed from that of commercial counterparts by <1 °C (figs. S19A-C), confirming that the heating mode preserves the solar-harvesting capability of conventional colored textiles rather than trading it away for cooling performance. In contrast, single-function white RC fabrics remain substantially colder under the same low-temperature conditions (Fig. 2H and fig. S20), underscoring the dual-mode advantage: one fabric combines the heating ability of commercial black fabrics with the cooling ability of white RC fabrics. This characteristic confers a distinct advantage over both conventional heating fabrics, which sacrifice color for functionality, and cooling fabrics, which sacrifice warmth. The SA fabric without the structural color coating also exhibited similar thermal management performance (figs. S20-S21), with comparable temperature regulation and radiative properties ($\rho_{solar}$ and $\varepsilon_{MIR}$). This confirms that the thermal regulation function of the CSA fabric is largely decoupled from the structural color, emphasizing that the core thermochromic switching *via* TCMs drives bidirectional thermal management, while the APC layer solely contributes to aesthetics. Theoretical calculations based on radiative heat transfer models(*27, 28*) (Fig. 4E) systematically quantified the cooling and heating power of CSA fabrics with distinct structural colors (blue, green, and red) under varying non-radiative heat exchange coefficients ($h$=0, 3, 6, 9, 12 W m$^{-2}$ K$^{-1}$), which represent different environmental convection conditions (from cool to hot). In a cool environment ($T_{amb}$=15 °C), the CSA fabrics achieved a net heating power density of ~300 W m$^{-2}$, accounting for solar absorption, thermal emission, and convective losses. In a hot environment ($T_{amb}$=35 °C), they delivered a net cooling power density of ~50-60 W m$^{-2}$. The phase transition temperature of TCMs governs this switchable power profile. This temperature-gated switching validates the autonomous transition of CSA fabrics from solar heating at low temperatures to radiative cooling at high temperatures, while maintaining stable performance across diverse climatic convection conditions.

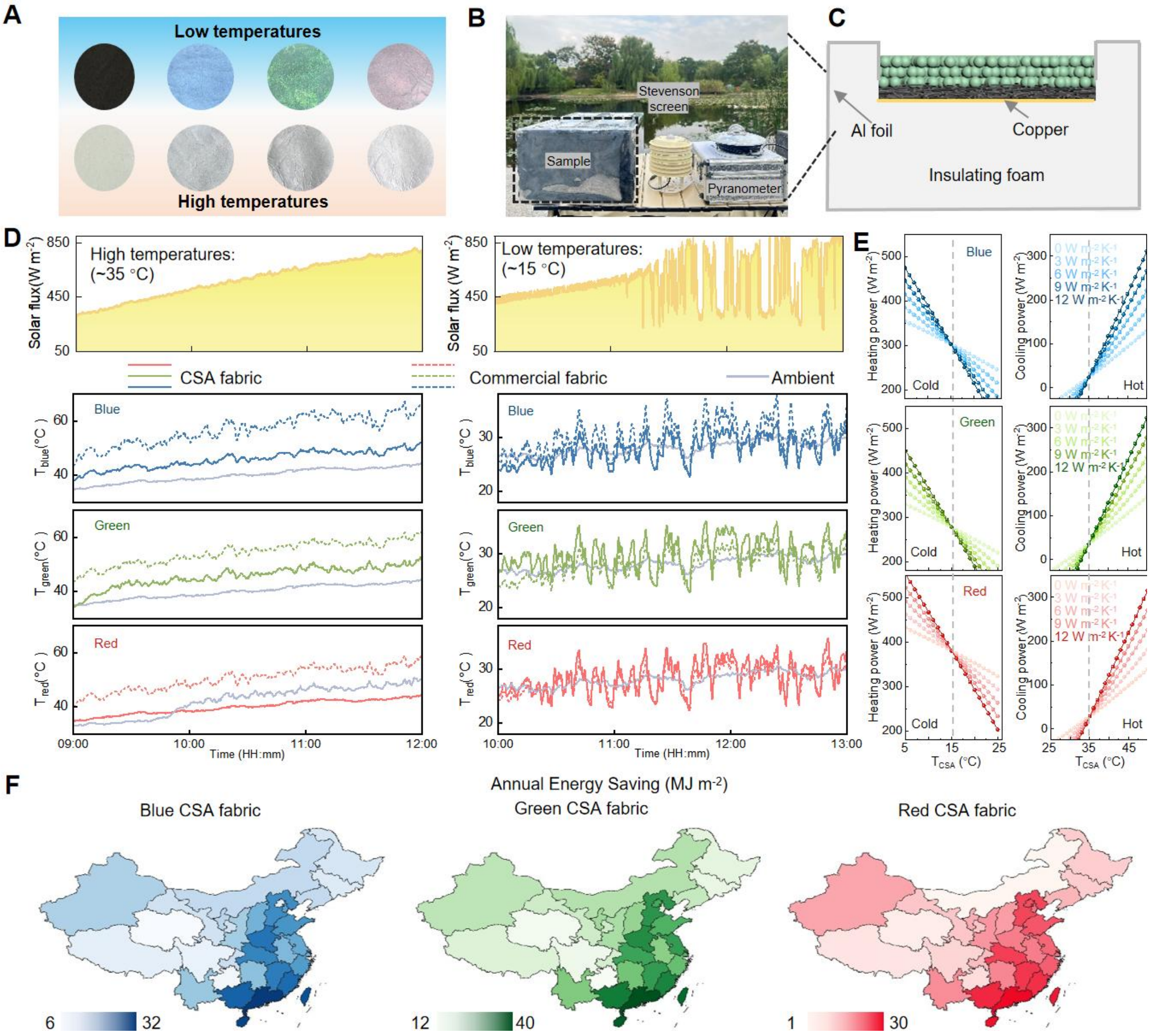

**Fig. 4.** Outdoor evaluation of dual-mode CSA fabric. (**A**) Photographic documentation of CSA fabrics under low (~15 °C) and high (~35 °C) temperature regimes. (**B-C**) Experimental setup for outdoor thermal regulation tests. (**D**) Outdoor temperature comparison between CSA fabrics and commercial-colored fabrics (left: high temperature; right: low temperature). (**E**) Theoretical calculations of cooling/heating power under varying convection conditions ($h$=0-12 W m$^{-2}$ K$^{-1}$). (**F**) Simulated annual energy-saving maps (China) of CSA fabrics.

To evaluate the energy-saving potential of the CSA fabric in practical building applications, we used EnergyPlus version 9.5 application(*29, 30*) to generate annual energy-saving maps for apartment buildings (fig. S22A), comparing CSA fabrics (blue, green, and red structural colors) with commercially colored fabrics. The analysis focused on dual-mode thermal regulation (heating/cooling) under real-world climatic conditions. As shown in the energy-saving maps (Fig. 4F (details in fig. S22B)), CSA fabrics of all colors demonstrated superior performance to commercial counterparts (fig. S22B), with energy savings of 5-15 MJ $m^{-2}$ across diverse climate zones. In Guangzhou (southern China, subtropical climate), the CSA fabrics were predicted to achieve an exceptional energy saving of ~40 MJ $m^{-2}$, which is significantly higher than the energy saving of commercial fabrics (fig. S22B). This advantage arises because colored commercial fabrics, with fixed solar absorptance, fail to meet cooling demands in high-temperature southern regions, leading to excessive HVAC (heating, ventilation and air conditioning) loads. In contrast, the CSA fabrics autonomously switch to a reflective white state ($T_{amb}$>35 °C) under intense sunlight, enhancing radiative cooling and reducing HVAC energy consumption compared to static commercial fabrics. Collectively, the CSA fabrics enable efficient dual-mode thermal management (cooling/heating) with color-agnostic performance and practical energy-saving potential.

**2.4 Application performance and multifunctional advantages of CSA fabrics**

A key demonstration of this design (Fig. 5A) is the CSA fabrics' color-rewritable behavior. A CSA fabric sample printed with a blue "Soochow" pattern underwent washing (anhydrous ethanol, 40 °C, 30 min) and drying (60 °C, 2 h), which effectively removed the original blue pattern. Subsequent spraying of a green "Zhang Group" pattern onto the cleaned surface transformed the fabric into a green-structured variant (Fig. 5B). Spectral analysis (fig. S23) confirmed that the rewritten pattern retained optical properties ($\rho_{solar}$) nearly identical to the original pattern of the same color series. This rewrite ability reflects a deliberate design principle that the aesthetic layer is treated as a reconfigurable interface rather than a permanent coating. Because the APC layer is held by physical self-assembly, it can be removed with a simple solvent wipe and re-sprayed in a new pattern on demand, decoupling appearance customization from the durability of the functional core, which is a capability inaccessible to dye-based coloration.

Beyond thermal and optical functions, the CSA fabric exhibits exceptional physicochemical properties derived from its PVDF-HFP matrix. The high fluorine content (C-F bonds)(*31*) endows it with hydrophobicity, evidenced by a water contact angle (WCA) of 132°±2° (Fig. 5C). Combined with its nanofiber structure, this property grants it excellent waterproofing and air permeability relative to commercial fabrics (Fig. 5D), which are critical for wearable comfort. These attributes position the fabric as a promising candidate for personal thermal management.

To comprehensively evaluate the potential of the CSA fabric for thermal management, we summarized and compared its performance with existing radiation-modulating materials(*13-15, 32*) in terms of functional decoupling, color versatility, thermal efficiency, wearability, and multifunctionality *via* a Radar plot in Fig. 5E. The modular architecture of CSA fabrics enables independent control over structural color, dynamic thermal switching, NIR reflection, and MIR emissivity. This design resolves the color-thermal trade-off, achieves full VIS spectrum coverage that surpasses the limited hues of prior work, and delivers a 30-40% higher solar reflectance contrast than static or single-mode fabrics, while integrating hydrophobicity and breathability for wearability. Unique color rewrite ability further adds multifunctionality. This comprehensive superiority showcases its potential as a next-generation thermal management fabric.

The real-world application can be vividly demonstrated through two prototypes. Fig. 5F shows a wearable smart thermal management fabric constructed by laminating a flower-patterned CSA fabric onto a dark gray garment. This assembly reversibly transitions to a white state at high temperatures while preserving its structural color at low temperatures. In a climate chamber (400 W $m^{-2}$ simulated sunlight), its surface temperature rose to 35 °C at 12 °C ambient (close to the garment's temperature) and dropped to 25 °C at 32 °C ambient (significantly lower than the garment), with thermal imaging (Fig. 5F inset) confirming localized heating/cooling. An intelligent tent ($20 \times 10 \times 10$ $cm^3$, Fig. 5G) fabricated from CSA fabric maintained a stable internal microenvironment: in a hot environment, the white state CSA fabric reflected solar radiation and dissipated heat *via* the atmospheric window; in a cold environment, the colored state (TCMs absorptive) absorbed solar energy while retaining hue, raising internal temperature. These prototypes underscore the fabric's adaptability for wearable and outdoor thermal management, leveraging autonomous color switching and dual-mode regulation.

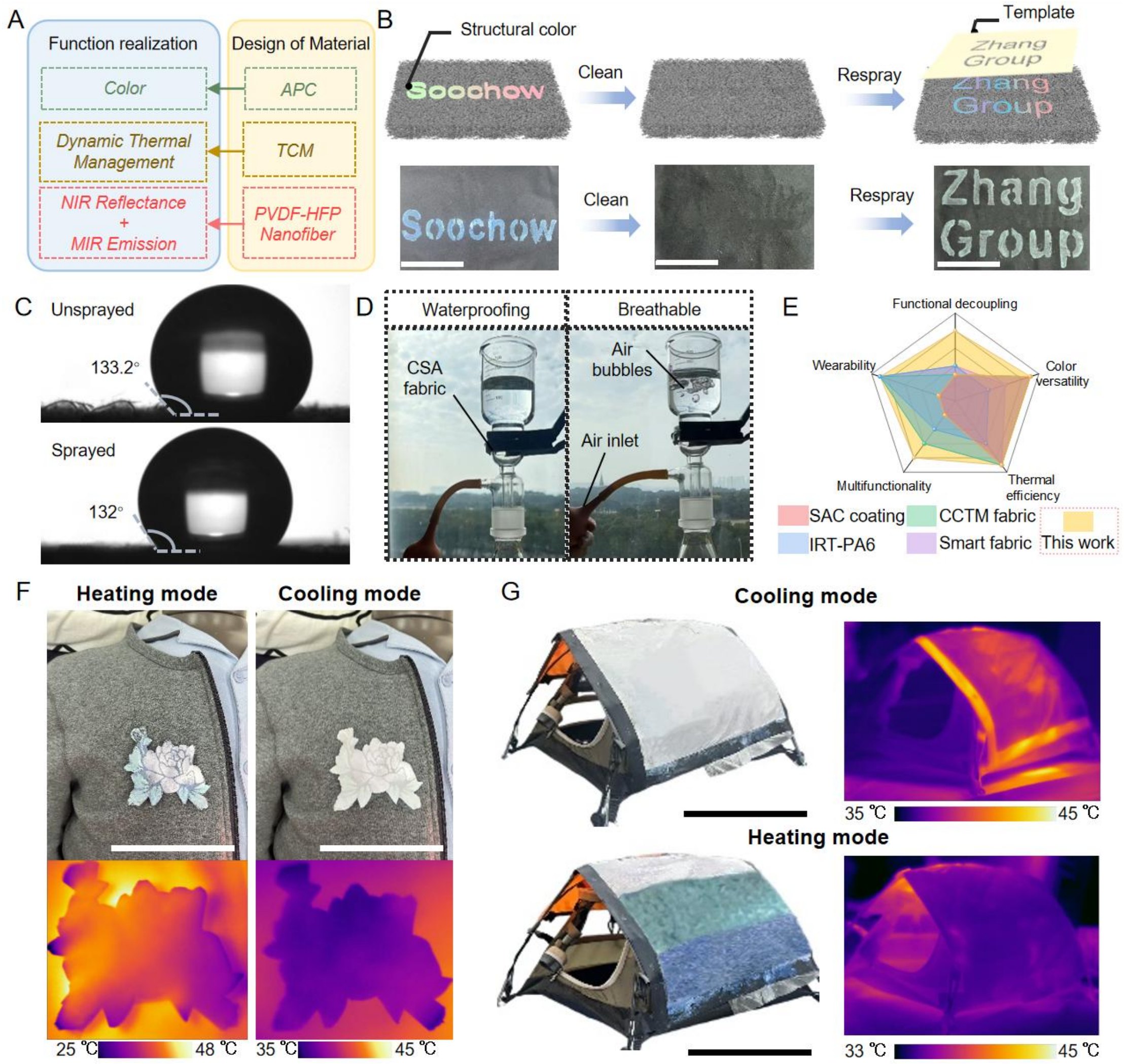

**Fig. 5.** Application performance and multifunctional advantages of CSA fabric. (**A**) Modular design of CSA fabric. (**B**) Color-rewritable behavior of CSA fabric (Scale bar: 4 cm). (**C**) WCA of CSA fabric. Both the SA fabric without sprayed PS nanoparticles and that with sprayed PS nanoparticles exhibited hydrophobicity. (**D**) Waterproofing and air permeability of CSA fabric. (**E**) Radar plot comparing CSA fabric with existing radiation-modulating materials (SAC: Superhydrophobic all-in-one cooling; CCTM: Carboxylated cellulose nanofibers/amorphous calcium carbonate/thermochromic micro-particles; IRT-PA6: Infrared (IR)-transparent-Nylon 6 (PA6) textile).(*13-15, 32*) (**F**) Wearable prototype (flower-patterned CSA fabric on garment) with thermal images (Scale bar: 10 cm). (**G**) Intelligent tent prototype maintaining stable internal

microenvironment (Scale bar: 10 cm).

## 3 Discussion

In summary, we introduce a CSA fabric that resolves the inherent conflict between aesthetic coloration and thermal management in personal thermal regulation. Through a modular design that segregates structural color from thermal function, the CSA fabric achieves autonomous bidirectional thermal regulation, performing radiative cooling under high-temperature conditions and passive heating under low-temperature conditions without compromising aesthetic versatility. This design breaks the limitation of traditional colored thermochromic materials, where color and thermal performance are in conflict, enabling color-agnostic thermal management across the VIS spectrum. The CSA fabric demonstrates remarkable versatility through its color-rewritable capability (pattern transformation *via* simple washing), exceptional physicochemical properties for wearability, and proven utility in wearable fabrics and intelligent tents. Energy simulations highlight its potential to reduce building HVAC energy consumption, emphasizing its value for sustainable thermal management. By merging aesthetic freedom with adaptive thermal intelligence, this work offers a concrete path to dynamic fabrics for personalized apparel and energy-smart architecture.

**Acknowledgements**

**Funding:** We appreciate the financial support of National Natural Science Foundation of China (No. 52303043), Jiangsu Provincial Science and Technology Program (Grant Number: BG2024020), Natural Science Foundation of Jiangsu Province of China (No. BK20211317, BK20220503), Major Basic Research Project of the Natural Science Foundation of the Jiangsu Higher Education Institutions (23KJA430014), China National Textile and Apparel Council Key Laboratory of Silk Functional Materials and Technology, and Key Laboratory of Jiangsu Province for Silk Engineering, the GRF grant from the Hong Kong Research Council (15204023).

**Author contributions:** Conceptualization: Y.W., Z.X., W.J., and J.H. Methodology: J.H. and K.-Q.Z. Investigation: Y.W., Z.X., W.J., and J.H. Visualization: T.-T.L. and H.C. Supervision: W.J., J.Z., H.W., and J.H. Writing—original draft: Z.X., Y.W., W.J., J.H., and K.-Q.Z. Writing—review & editing: Z.X., Y.W., W.J., J.H., S.S., J.F. and K.-Q.Z. Funding acquisition: H.W., J.F., J.H. and K.-Q.Z.

**Competing interests:** The authors declare that they have no competing interests.

**Data and materials availability:** Materials and methods are provided in the Supplementary Materials. All data needed to evaluate the conclusions in the paper are present in the paper and/or the Supplementary Materials.